\newcommand\pd[2]{\frac{\partial#1}{\partial #2}}
\newcommand\ppdd[3]{\frac{\partial^2#1}{\partial #2\partial#3}}

\newcommand\cX{{\cal X}}

\newcommand\tm{\(T(M)\)}
\newcommand\tsm{\(T^*(M)\)}
\newcommand\ttsm{\(TT^*(M)\)}
\newcommand\ttm{\(TT(M)\)}
\newcommand\tstm{\(T^*T(M)\)}
\newcommand\tstsm{\(T^*T^*(M)\)}

\newcommand\tbh{H_T}

\newcommand\ds{\displaystyle}

\documentclass{article}
\usepackage{amsfonts}
\usepackage{graphicx}

\title{Free Lunch}
\author{George Svetlichny\footnote{Departamento de Matem\'atica, Pontif\'{\i}cia Universidade Cat\'olica, Rio de Janeiro, Brazil \newline
svetlich@mat.puc-rio.br \hfill \url{http://www.mat.puc-rio.br/\~svetlich}}}
\begin{document}
\maketitle
\begin{abstract}
The Free Lunch Principle: Nature thrives on freebies. She chooses nothing, and no one helps Her. She must use canonical mathematical structures as there is no one to tell Her otherwise. With this I show where variational principles are superfluous and Noether's theorem is trivial. This paper is based on a talk I gave at the Harvey Brown Festschrift in Oxford, July 2015.
\end{abstract}

\section{Introduction}\label{sec:intro}

There ain't no such thing as a free lunch. This is an oft quoted fact of life but it just ain't so in
mathematics. By a free lunch I mean a mathematical structure that exists
without external specification. These are among what mathematicians call  \emph{canonical structures} and
these arise from the \emph{mere} construction of other structures. I suggest that it is these structures that Nature uses.
For simplicity, my examples will be taken from classical particle mechanics. In what follows I use standard notation and constructs from
manifold theory, essentially vector fields, differential forms, and various bundles. Readers unfamiliar with these should consult
subsection \ref{man} of the Appendix for a quick tour of the needed material. Free lunches are just ain't visible without this perpective.

We shall deal with a manifold \(M\) with local coordinates \(q^i,\, i=1,\dots,n\) representing  the configuration space  (positions of the particles).\footnote{There may  be no global system of coordinates.} Two immediate objects of interest are the tangent bundle \tm\ (where Lagrangians live) and the cotangent bundle \tsm, known as\emph{ phase space} (where Hamiltonians live).

An important example of a free luch is the  canonical 1-form on phase space \(\theta = p_idq^i\) (summation convention in force).
Usually to have a one-form
on a manifold you have to specify it externally for your own obscure reasons.
On the cotangent bundle \emph{it's just there}. It arises from the very definition of the
cotangent bundle. The canonical \(1\)-form then defines the symplectic form \(d\theta= dp_i\wedge dq^i\)
along with the family of Hamiltonian dynamical systems. This may account for the prevalence of Hamiltonian systems
in nature. It's a class of dynamical systems free for the taking. This family is a quintessential free lunch. More on this later.

Why is \(\theta\) a free lunch?  A \(1\)-form on \(M\)  is written as \(\alpha_jdq^j\). The pairs \((q^i,\alpha_j)\) are coordinate functions on \tsm. Physicists usually write the coordinates as \((q^i,p_j)\) and we shall do so from now on. Note that this pair represents a point \(q\) in \(M\)  and a \(1\)-form \(p_jdq^j\) \emph{on \(M\)  at the given point} \(q\).   Shortly we shall use the expression \(p_jdq^j\) to mean the \(1\)-form \(\theta\) \emph{on \tsm\ at the point \((q^i,p_j)\)}. Such double meanings of expressions is common among free lunches. To define \(\theta\) at the point  \((q^i,p_j)\) we have to say how it acts (contracts) on a vector at the same point. Let \(\displaystyle \cX=\eta^i\pd{}{q^i}+\sigma_j\pd{}{p_j}\) be such a vector. Define \(\langle\theta,\cX\rangle=p_i\eta^i\). Notice that in this definition one just uses structures already present and no further objects need be introduced by the mathematician, or physicist, or anybody else. See subsection \ref{aptheta} of the Appendix for a geometric depiction of this argument.

There are many other free lunches such as the
canonical map  \(j:V\to V''\) between a vector space \(V\) and its double dual \(V''\) given by
\(\langle j(v),\phi\rangle=\langle \phi,x\rangle\). Note the double meaning of \(\phi\) in this expression, as a point in a vector space and as a linear form.

Another free lunch is the Lie-Jordan algebra associated to an associative algebra \({\cal A}\). There is the
Lie product: \([a,b]=ab-ba\) and the Jordan product: \(a\ast b=ab+ba\). Every associative algebra is a Lie-Jordan algebra, extra structure free to use. Is it a surprise that quantum mechanics uses them?

Akin to a free lunch is the central limit theorem of probability. Many independent
stochastic influences lead to the universal existence of the Gaussian
distribution. This may have something to do with physical laws being beautiful
and at most second order. As the universe is made up of very many parts, all
ugliness and higher order contributions just get squeezed out by something like
central limit theorem. Taking this a bit further, if any behavior of constituent
parts, when there are very many of them, leads to a universal pattern one could take the attitude
that such a universal pattern can exist for no reason at all. It's just a pattern
there for the taking, and Nature takes it. If \emph{anything} leads to a pattern then
\emph{just nothing itself} leads to it.
Maybe there are no fundamental building blocks to the world. Nature could just be manifesting the universal
patterns that are just there.

Free lunches need no reason to be used. There is no deity or some ``fundamental
principle" or ``law" needed to put them into practice. Just take and enjoy.
So if Nature is consistently doing something we can't quite understand, the
question to ask is ``where's the free lunch?" If there is some central kitchen
cooking up free lunches, then, if we can find it, we shall have a theory of everything as Nature
would be feasting there.

I won't try to define free lunch with any rigor, but just paraphrase
the American Supreme Court Justice Potter Stewart:
``I can't define free lunch, \emph{but I know it when I see it!"}\footnote{For the curious, the Honorable Justice Stewart was referring to pornography.}

\section{Where variational principles are not needed}

I was always bothered by variational principles in physics. Why should Nature
care about extremizing an action integral
\[S(\phi) = \int L(x, \phi(x), \partial_\mu\phi(x))\, d^nx? \]
I of course don't question this when one is really trying to minimize length or
area or energy or something like that, but why care about the action integral
for the Yang-Mills gauge theory coupled to fermions? What is Nature looking
for? I decided  She doesn't care about the integral. Besides, the
integral may not exist for the fields one is interested in. Most of the variational
calculations done by theoretical physicists are formal in any case, so they too
don't care about the integral as such. What is going on?
\footnote{As a bothersome aside, the action integral seems to acquire importance in
quantization using Feynman's integrals, but that's another kettle of fish. Where's the free lunch?! Here I'll be
classical for now.}

If the action integral is a decoy,
there should be a way of understanding all the benefits of ``variation" without
recourse to it. Where's the free lunch? For a starter, consider just systems
with a finite number of degrees of freedom and first order Lagrangians \(L(q^i, v^j)\)
where \(v\) is velocity.\footnote{What velocity \emph{really} means we'll see later.}
Variational calculus supplies us with a second-order ordinary
differential equation starting with \(L\) and it's first and second order derivatives.
We must imagine how this can proceed in a coordinate free fashion. Taking first order
derivatives means essentially computing \(dL\)
which thus lives (is a section of) on
\(T^*(TM)\). There must yet be another step to get to second order derivatives,
which we won't need here. Where do second-order ordinary differential equations live?
They are special sections of
\ttm, to be explained below. So how does one get from a section of \tstm\
and wherever the second order derivatives live to a special section of \ttm\ in
a coordinate free fashion and universally for any \(L\) that you may choose? This
is only possible if there are canonical relations that are just there, among the
various bundles.

The intrinsic mathematical description of  variation calculus involves iterated tangent
and cotangent bundles. The four \ttm, \ttsm, \tstm, \tstsm\ cover
most of the situations for first order Lagrangians. These four double bundles
offer a modest free banquet.

My previous attempt at getting Euler's equations without the variational principle can be found in \cite{sveeq} where equivariance was the guiding principle. I did not get Euler's equation uniquely from equivariance, but a narrow class of possible equations. Below I show a much quicker way based only on dimensional analysis and free snacks in the iterated bundle.

First some elementary mechanics.

 Consider the kinetic energy:
\[\hbox{KE}=\frac12mv^2=\frac1{2m}p^2.\]
One can't square a vector or co-vector\footnote{One needs a metric, but here is no free metric on \tm\ or \tsm.}; but\ldots \(\ds\frac12mv^2=\frac12(mv)v=\frac12pv\). This is a perfectly legitimate contraction, a vector with a co-vector.

I will state Newton's Laws as:
\begin{equation}\label{Newton}
v=\frac {dq}{dt},\quad f=\frac {dp}{dt}
\end{equation}

Newton did not consider the first equation as a physical law but as an expression of  velocity. As one of the creators of calculus, he would not object to the equation. I consider this equation as a physical law. Velocity is an attribute that a particle (or system) has instantaneously in contrast to Zeno's idea that it would be instantaneously at rest.\footnote{Zeno was an earthling (as far as I know) and if he had known general relativity then he would have realized that with the earth present, the space time metric around a flying arrow is essentially different from that around an arrow at rest. Thus space-time knows the difference and Zeno should have. The first equation is an approximation to the true physical situation.}  Thus the first equation is  a physical law on par with the second.

We need to consider \(p\) and \(v\) together. Now \(v\) lives in \tm, \(p\) in \tsm.\footnote{That we can write KE\(\ds=\frac12 pv\) shows that mass establishes a relation betwen \tm\ and \tsm, a fact not yet explored in the literature.}
To see them together try \tstm\ which has elements
\begin{equation}\label{tstm}
 (q^i,v^i;A_idq^i+B_idv^i).
\end{equation}
Denote these by \((q,v;A,B)\). As was pointed out above, \(v\) here is considered as a physical entity (Zeno's oversight). Mathematicians don't think like this, but physicists should.

Under change of coordinates \(q\mapsto \tilde q(q)\) one has:

\[\tilde A_i=A_j\frac{\partial q^j}{\partial \tilde q^i}
+ B_j \tilde v^k \frac{\partial^2 q^j} {\partial
\tilde q^k\partial \tilde q^i},\quad \tilde B_i=B_j\frac{\partial q^j} {\partial \tilde
q^i}\]
\(B\) is a co-vector identified with momentum, (writing it as \(p\) would be more appropriate), but what is \(A\)?

It turns out that \(A\) is force. There are various ways to see this, we present two:
\begin{enumerate}
  \item The free lunch way: \(\theta=p_i\,dq^i\) is free lunch, take time derivative: \(\dot\theta=f_i\,dq^i+p_i\,dv^i\). Compare with (\ref{tstm}).
 \item A more physical way: \(p_iv^i\) is a canonical scalar in  \tstm\ so is its time derivative. Thus \(\tilde f_i\tilde v^i+\tilde p_i\tilde a^i=f_i v^i+\ p_i a^i\). All transformations other then for \(f\) are known, and we deduce \( \ds\tilde f_i = f_a\pd{q^a}{\tilde q_i} + p_j\ppdd{q^j}{\tilde q^b}{\tilde q^i}\tilde v^b\), same as for \(A\).
\end{enumerate}

We can now give a variationless derivation of Euler's equation. We have identified the physical dimensions of the components of \tstm:
\[(q,v;A,B)=(\hbox{position},\hbox{velocity}; \hbox{force}, \hbox{momentum})\]
Consider now a Lagrangian \(L(q,v)\)  and take its differential:
 \[\quad dL=\pd L{q^i}\,dq^i+ \pd L{v^i}\,dv^i\]
So we now conclude: \(\displaystyle \pd L{q^i}\) is force and \(\displaystyle \pd L{v^i}\) is momentum

By Newton's laws (\ref{Newton}) we conclude:
\begin{equation}\label{euler}
  v^i=\frac{dq^i}{dt}, \quad \pd L{q^i}=\frac d{dt}\pd L{v^i}
\end{equation}

The second equation is Euler's equations, showing it as being a free lunch. No need for a variational principle.

Other iterated bundles are also of interest. We have:

\[TT^*(M):\quad\left(q^i, p_i, S^i\frac{\partial}{\partial q^i}  +
T_j\frac{\partial}{\partial p_j}\right)=(q,p;S,T)\]
 \[T^*T^*(M):\quad\left(q^i, p_i,Q_i\, dq^i  +
R^j\, dp_j\right)=(q,p;Q,R)\]
\[TT(M):\quad \left(q^i, v^i, U^i\frac{\partial}{\partial q^i}+
V^j\frac{\partial}{\partial v^j}\right)=(q,v;U,V)\]

One has the following canonical isomorphisms\footnote{``Canonical" here means the isomorphisms commute with coordinate changes on \(M\).} (free lunches):
\begin{equation}\label{to1}
  T^*T(M)\to TT^*(M):\quad (q,v;A,B)\mapsto (q,B;v,A)
\end{equation}
\begin{equation}\label{to2}
  T^*T^*(M)\to TT^*(M):\quad (q, p; Q, R)\mapsto (q,p;R,-Q)
\end{equation}

Thus the physical dimensions are:
\vskip 5pt
\centerline{In \tstm: (position, velocity; force, momentum)}
\vskip 5pt
\centerline{In \ttsm: (position, momentum; velocity, force)}
\vskip 5pt
\centerline{\hskip 1em In \tstsm: (position, momentum; -force, velocity)}
\vskip 5pt
The bundle \ttm\ apparently has no canonical relation to the other three bundles.\footnote{Because of (\ref{to1}) and (\ref{to2}) any two iterated bundles such as \(T^*TTT^*\cdots T^*TM\) having the same number of iterations and at least one \(T^*\) functor applied, are canonically isomorphic. The bundle \(TTTT\cdots TTM\) with purely the \(T\) functor applied, stands apart.}
 It's physical role is very different and not at all clear. The physical dimensions of its components are (position, velocity; \emph{velocity}, acceleration). There are two velocities, the first one, the above referred to physical attribute, the second one, the time derivative of position. The acceleration should also be considered as a physical entity being the time derivative of the physical velocity. This bundle carries some indication that the first of Newton's laws (\ref{Newton}) is a true physical law and not a definition. I'll come back to this bundle in the next section.

Consider now \(H(q,p)\), the hamiltonian function in \tsm. One has the differential
\[dH=\pd H{q^i}\,dq^i+\pd H{p_i}\,dp_i.\]
The physical dimensions in \tstsm\ are: (position, momentum; -force, velocity), thus:
\[\pd H{q^i} \hbox{\ is -force, and\ } \pd H{p_i} \hbox{\ is velocity.}\]
By Newton's laws  (\ref{Newton}):
\[\frac {dq^i}{dt}=\pd H{p_i},\quad\frac {dp_i}{dt} =-\pd H{q^i}\]
which are precisely Hamilton's equations.
\section{Hamilton is cleaner}

It is known that the two schemes, Euler and Hamilton, are equivalent\footnote{Under a condition to be stated below} and there is a geometric way of seeing this.

The graph of  \(dH\) as a subset of \tstsm\ is the following set..

\begin{equation}\label{hinl}
\left.\left\{\left(q^i, p_i, \pd H{q^i}; \pd H{p_i}\right)\,\right|\,(q,p)\in T^*(M) \right\}.
\end{equation}
We now use the canonical equivalence \tstsm\(\simeq\)\tstm\ to bring this sets to \tstm.
\begin{equation}
  \left.\left\{\left(q^i, \pd H{p_i}; -\pd H{q^i}, p_i\right)\,\right|\,(q,p)\in T^*(M) \right\}.
\end{equation}
Since differentials of functions in \tm\ live in \tstm, this should be  the graph of \(dL\) for some function \(L\) which physicists call the Lagrangian.  This means that
\begin{equation}\label{htol}
 p_i=\pd L{v^i}\left(q,\pd H{p}\right),\quad-\pd H{q^i} =\pd L{q^i}\left(q,\pd H{p}\right)
\end{equation}
These are well know relations from which  \(L\) can be reconstructed from \(H\) as
\[L\left(q,\pd H{p}\right)=p_i\pd H{p_i}-H\]
To express this in \((q,v)\) coordinates one must solve \(\ds v^i=\pd H{p_i}\) for \(p\) as a function of \(v\) and this can be done if \(\ds (q, p)\mapsto \left(q, \pd H{p}\right)\) is invertible and we assume this from now on. Equivalently the map \(\ds (q,v)\mapsto \left(q, \pd L{v}\right)\) is also invertible and it is under these conditions that Euler and Hamilton are equivalent.

A geometric way of seeing this is to note that we have switched two ``axes" (second and fourth) which depict the graph of the \(p\) \emph{derivative} of \(H\). A sign was changed also.

Switching axes on graphs of derivatives is a well known procedure known as the  Legendre transform.

\begin{center}
\begin{picture}(250,100)(0,0)
\put(0,0){\line(0,1){100}}
\put(0,0){\line(1,0){100}}
\qbezier(0,0)(40,100)(100,100)
\put(50,50){\makebox(0,0){\(f'(x)\)}}
\put(125,60){\makebox(0,0){Axes}}
\put(125,50){\makebox(0,0){\(\mapsto\)}}
\put(125,40){\makebox(0,0){switch}}
\put(150,0){\line(0,1){100}}
\put(150,0){\line(1,0){100}}
\qbezier(150,0)(250,40)(250,100)
\put(200,50){\makebox(0,0){\(g'(x)\)}}
\end{picture}
\end{center}

Now \(g\), the Legendre transform of \(f\), is defined by: \(g(x)=x(f')^{-1}(x)-f((f')^{-1}(x))\).
A possibly more familiar form is \(g(f'(x))=xf'(x)-f(x)\). This can be easily shown to correspond to switching axes on graphs of derivatives. As before we are ignoring additive constants.

The equivalence \ttsm\(\simeq\)\tstsm\ exchanges vectors and \(1\)-forms. The vector corresponding to
\(\theta =p_i\,dq^i\) is \(\ds\Theta=p_i\pd{}{p_i}\).

Now on any vector space \(V\), with linear coordinates \(x^i\),
\[\Theta_V=x^i\pd{}{x^i}\]
is a canonical vector field (free lunch).

The Legendre transforms \(L\mapsto \Theta_{TM}L-L=H\) and \(H\mapsto \Theta_{T^*M}H-H=L\) makes use of this free lunch.\footnote{In the first equation \(H\) is in \((q,v)\) coordinates and in the second equation \(L\) is in \((q,p)\) coordinates. In fairness, one could add constants to these transforms which does not change anything essential. If we want to include gravity, then the absolute scale of energy is important and constants need to be considered.}

A first order ordinary differential equation \(\dot q^i=\xi^i(q)\) is a vector field \(\cX\) on \(M\) given by  \(\ds\cX=\xi^i\pd{}{q^i}\). A second orger differential equation should by all rights be a vector field on \tm, but it's not any vector field that defines such an equation. Consider the equation
\(\ddot q^i=\xi^i(q,\dot q)\). To associate a vector field to it we turn it into a first order system:
\begin{eqnarray*}
  \dot q^i  &=& v^i \\
  \dot v^i &=& \xi^i(q,v)
\end{eqnarray*}
which corresponds to the vector field which at a point \((q,v)\) is  \(\ds v^i\pd{}{q^i}+\xi^i\pd{}{v^i}\). Note that this is on that part of the bundle where \emph{velocity} \(=\) velocity, that is where the first of Newton's laws (\ref{Newton}) holds. The vector field \(\cX\) on \tm\ that correspond to Euler's equation  (second equation of (\ref{euler})) is easily shown to be determined by:
\begin{equation}\label{eu}
\cX\rfloor d\theta_T + d\tbh{}=0,
\end{equation}
 where
\(\tbh=\Theta_{TM}L-L\) is the Hamiltonian function in \((q,v)\) coordinates in \tm\ and
\(\ds\theta_T=\pd L{v^i}\,dq^i\), one of the equivariant \(1\)-forms\cite{sveeq} in \(TM\) and the pullback of the canonical (free lunch) \(\theta\)
in \tsm\ by \(\ds(q,v)\mapsto \left(q,\pd L{v}\right)\).

Thus to define the 2\({}^{nd}\) order differential equation that is Euler's, indirect references have to be made to structures in \tsm\ (phase space), a different bundle.

Compare (\ref{eu}) with the equation defining the Hamitonian field in \tsm:
\begin{equation}\label{ha}
\cX\rfloor d\theta + dH=0.
\end{equation}
In contrast to (\ref{eu}), this equation uses the free-lunch canonical \(\theta\) and no indirect references to another bundle. One might say that Euler's equation is Hamilton's equation seen through a glass darkly. It also hints that on the Hamiltonian side, in \tsm, many constructs could be clearer and more natural, that is, free lunches, as we shall see below.

We now show the equivalence of Hamilton's and Euler's solutions.

Let \((q(t),p(t))\) be a path in \tsm\ satisfying Hamiltons equations. Lifted to \ttsm\ it becomes
\begin{equation}\label{hpath}
(q(t),p(t);\dot q(t),\dot p(t))
\end{equation}
The Hamitonian vector field (section of \ttsm) has the form:
\begin{equation}\label{hinla}
\left(q, p; \pd H{p},-\pd H{q}\right)
\end{equation}

As before, using a canonical isomorphism we transfer (\ref{hpath}) and (\ref{hinla})  to \tstm\ (where \(dL\) lives):
\begin{eqnarray}\label{pp}
&&(q(t),\dot q(t);\dot p(t), p(t)) \\
 &&\left(q,\pd H{p} ; -\pd H{q},p\right)
\end{eqnarray}
The latter is the graph of \(dL\), that is:
\[\left(q,\pd H{p} ; \pd L{q}\left(q,\pd Hp\right),\pd L{v}\left(q,\pd Hp\right)\right)\]
But \(\ds\pd Hp = \dot q\) so  on the path this is:
\begin{equation}\label{ppinvp}
\left(q(t),\dot q(t) ; \pd L{q}\left(q(t),\dot q(t)\right),\pd L{v}\left(q(t),\dot q(t)\right)\right)
\end{equation}
Comparing entries in  (\ref{pp}) and (\ref{ppinvp}) one has on the given path:
\[v=\frac{dq}{dt}, \quad \pd L{q}=\frac d{dt}\pd L{v}\]
That is, the path satisfies Euler's equation.

The argument can be run backwards to go from a solution of Euler's equation to that of Hamilton's.

Free lunch may not really explain why Lagrangian theories are so successful,
but it does shed a light. So we all shove stuff into the physicist's machine and get stuff
back. We say, ``Wow! We can get General Relativity, we can get the Standard
Model, we can get String Theory, we can Tame Ferocious Tigers, aren't we
clever!" Nature doesn't care we do this, we are her naughty kids; besides the
factory is free to use by anyone. Exhilarating stuff, but let's look around and
ask (1) What are Nature's machines made of; how do they work? (2) What is
She banging together in Her hangar.
Every time I have ever asked the first question the answer has been: Nature
builds from parts that are just there and are free for the taking. What else
could She use? There are no suppliers. I wish I knew the answer to the second
question.

\section{Where Noether's theorem is trivial}

First I review the usual variational argument that leads to Noether's theorem. Let \(L(q,v)\) be a Lagrangian and consider\footnote{Why would you want to do this?} the integral:
\begin{equation}\label{intL}
  \int_{t_1}^{t_2}L(q(t),\dot q(t))\,dt.
\end{equation}
Consider now a ``variation" of the \(q\) coordinates \(\tilde q^i=q^i+\epsilon \xi^i\) where \(\epsilon\) is considered infinitesimal\footnote{The physics literature often writes \(\epsilon \xi^i\) as \(\delta q^i\) but this practice obscures even further what is really going on.} and \(\xi^i\) are functions of \(q\).\footnote{We are dealing here with \emph{point symmetries}. More complicated ``variations" can be considered, but  these already make the main point.}
The assumption that is now made is the vanishing of  the ``variation" of (\ref{intL}) meaning the vanishing of the linear term in \(\epsilon\) in a Taylor expansion of:
\[\int_{t_1}^{t_2}L(\tilde q(t),\dot {\tilde q}(t))\,dt.\]
A simple chain rule calculation reveals that the ``variation" is:
\begin{equation}\label{vint0}
\int_{t_1}^{t_2}\left(\pd{L}{q^i}( q(t),\dot q(t))\xi^i(q(t))+\pd{L}{v^i}( q(t),\dot q(t))\pd{\xi^i}{q^j}(q(t))\dot q^j(t) \right)\,dt.
\end{equation}
Now the hypothesis is that this must vanish identically for all paths \(q(t)\). Thus the integrand must vanish on all paths but the physics literature does not take this step and using integration by parts rewrites\footnote{I'm leaving out the arguments \(q(t)\) and \(\dot q(t)\) of the functions involved} the integral as:
\[\int_{t_1}^{t_2}\left(\left(\pd{L}{q^i}-\frac d{dt}\pd{L}{v^i}\right)\xi^i+\frac d{dt}\left(\pd{L}{v^i}\xi^i\right)\right)\,dt.\]
The identical vanishing of the integrand is written thus:
\[\left(\pd{L}{q^i}-\frac d{dt}\pd{L}{v^i}\right)\xi^i+\frac d{dt}\left(\pd{L}{v^i}\xi^i\right)=0.\]
Now comes the leap of the quantum cat: \emph{on a path that satisfies Euler's equations} the first term vanishes, so on solutions of Euler's equations the second term vanishes, and the expression \(\ds \pd{L}{v^i}\xi^i\) is a constant. \emph{Under vanishing of the ``variation" of (\ref{intL}) solutions of Euler's equation have an associated constant of motion.} This is Noether's theorem which is obviously an important insight but the above deduction is strangely mysterious. Physics literature just presents the steps without any justification or insight beyond it just working.

Now to the simplification. The ``variation" of the coordinates is just an obscure way of talking about a vector field:
\(\ds \cX=\xi^i\pd{}{q^i}\). Lagrangians live on \tm\, and the vector field \(\cX\) has a canonical lifting\footnote{See subsection \ref{lvf} of the Appendix.} \(\cX_{T}\) to \tm\ given by
\[X_T=\xi^k\pd{}{q^k}+v^j\pd {\xi^k}{q^j}\pd {}{v^k}.\]
The identical vanishing of the integrand in (\ref{vint0}) for all paths \(q(t)\) is now simply seen as the statement:
\begin{equation}\label{xtL}
\cX_T(L)=0,
\end{equation}
which is a direct statement of invariance, the Lagrangian is constant on the integral paths of the vector field. One could just take this as the initial assumption of symmetry without invoking the integral and going through all those manipulations. One can rewrite (\ref{xtL}) to get Noether's theorem just as before,\footnote{There is a canonical way of doing this but this would take us too far afield, and besides, the free lunch here is not very tasty.} but the situation is even simpler if we consider  Noether's theorem  in \tsm\ and I do so now.

The vector field \(\ds\cX\)  in \(M\) also lifts canonically\footnote{See subsection \ref{lvf} of the Appendix.} to \tsm\ as:
\[\ds\cX_{T^*}=\xi^i\pd{}{q^i}-p_j\pd{\xi^j}{q^i}\pd{}{p_i}.\]
Invariance of the Hamitonian, i.e. symmetry of dynamics is given by {\(\cX_{T^*}(H)=0\).
Due to the equivalence \(T^*T^*(M)\simeq TT^*(M)\), the associated \(1\)-form to \(\cX_{T^*}\) is: \(\ds p_j\pd{\xi^j}{q^i}\,dq^i+\xi^i\,dp_i\). This form is exact, equal to \(dK_{\cX}=d(p_i\xi^i)\).
Thus the symmetry flow is Hamiltonian, and
\(\cX_{T^*}(H)=0\) now becomes: {\(\{K_{\cX},H\}=0\).\footnote{See subsection \ref{Poiss} of the Appendix.}

This is Noether's Theorem! Read one way -- deformation of \(H\) generated by \(K_{\cX}\) is zero, this is a statement of symmetry.
Read the  other way -- time derivative of \(K_{\cX}\) under time evolution is zero, this is a statement of conservation.

The conserved quantity is \(p_i\xi^i\) which in \((q,v)\) coordinates is \(\ds\xi^i\pd L{v^i}\) just as in the variational case.

Rather than  ``a profound connection between symmetries and conservation" as is constantly stated in physics books, Noether's theorem here is just reading a vanishing Poisson bracket in two separate ways. Getting the theorem through variational calculus gives an impression of profundity, but that's an illusion. The only bit of mathematics here is recognizing that the lifted field to \tsm\ is Hamiltonian, which is rather immediate.

The above view was presaged by a question posed by Dwight E. Neuenschwander in  American Journal of Physics, \textbf{63}, 489 (1995). Parapharasing:
``Is there a Noether's theorem for discrete symmetries?''
Two opposite answers were given in American Journal of Physics, \textbf{64}, 849 (1996)). The ``no" answer was given by Benito Hern\'andez-Bermejo, and the ``yes" answer for quantum theory by Robert Mills who wrote:

 ``\emph{The observable \(A\) is invariant under the transformation generated by \(\hat B\) if and only if the observable \(B\) is invariant under the
 transformation generated by \(\hat A\).''}
 This is reading \([\hat A,\hat B]=0\) in two ways.

Leaving the question of discrete symmetries aside, this is precisely the argument I gave above for classical theories.

So we have Noether's theorem in \tm\ and \tsm. How do they compare?
One has the following theorem:
\[\cX_{T^*}(H)=0 \Leftrightarrow \cX_{T}(L)=0\]
Thus Lagrangian and Hamiltonian point symmetries are the same. The proof is straightforward but tedious. See the subsection \ref{eueha} in the Appendix.

\section*{Appendix}

\subsection{Quick tour of relevant manifold formalism}\label{man}

The configuration space (positions of the particles) will be a manifold \(M\) with local coordinates \(q^i,\, i=1,\dots,n\).\footnote{There may not be a global system of coordinates.} The particles would have velocities \(v^i\) and joining both coordinates as \((q, v)\) creates a new manifold called the \emph{tangent bundle} of \(M\), denoted by \tm. I will use the mathematician's designation for vectors: \(\ds v= v^i\pd{}{q^i}\), where I have adopted the summation convention in that a repeated upper and lower index is summed over.\footnote{Without this convention a mathematician would write \(\ds v=\sum_iv^i\pd{}{q^i}\).} A vector is a differential operator and given a vector \(\ds \cX=\xi^i\pd{}{q^i}\) and a function \(f\) of the variables \(q\) one has \(\ds \cX f= \xi^i\pd{f}{q^i}\).

A \emph{differential \(1\)-form} is a linear form on vectors (also know as a \emph{co-vector}, or \emph{covariant vector}) and is usually written with subscript indices: \(\alpha_i\). The form \(\alpha\) applied to the vector \(v\) gives the number \(\langle \alpha, v\rangle = \alpha_iv^i\) known as the \emph{contraction} of \(\alpha\) with \(v\). The mathematician's designation for a one-form is \(\alpha =\alpha_i\,dq^i\). The form \(dq^i\) is a special case of converting a function \(f\) of the variables \(q\) to a one form known as the differential of \(f\)  by defining \(\ds df=\pd{f}{q^i}\,dq^i\). In physics, momenta are \(1\)-forms.\footnote{In practically all elementary physics books there is no distinction between vectors and \(1\)-forms as the metric in \(3\)-space can be used to convert one to the other. This completely obscures their truly different \emph{physical} nature. Physicists, and especially their students, should know this.} Joining the coordinate \(q^i\) with the coefficients of a \(1\)-forms \(\alpha\) as \((q,\alpha)\) creates a new manifold called the \emph{cotangent bundle} of \(M\), denoted by \tsm. Physicists know \tsm\ as \emph{phase space}.

There are higher order forms, a \(k\)-form is a \emph{totally antisymmetric} covariant \(k\)-tensor \(\omega_{i_1i_2\cdots i_k}\). Mathematicians write this as
\[\omega =\omega_{i_1i_2\cdots i_k}\,dq^{i_1}\wedge dq^{i_2}\wedge\cdots\wedge dq^{i_k}.\]
The number \(k\) is known as the order of the form and we write \(k=|\omega|\). A \(0\)-form is just a function, and on a manifold of dimension \(n\) there are no \(k\)-forms for \(k>n\).
The product \(\wedge\) is bilinear, distributes over sums, and satisfies \(\alpha\wedge\beta=(-1)^{|\alpha||\beta|}\beta\wedge\alpha\). There is a differential operator \(d\) called exterior derivative defined by \(d\omega =d\omega_{i_1i_2\cdots i_k}\wedge dq^{i_1}\wedge dq^{i_2}\wedge\cdots\wedge dq^{i_k}\). This is a \((k+1)\)-form. One has the Leibnitz rule \(d(\alpha\wedge \beta)=d\alpha\wedge \beta + (-1)^{|\alpha|}\alpha\wedge d\beta\), furthermore \(d^2=0\). A \(k\)-form \(\alpha\) is \emph{exact} if \(\alpha=d\beta\) for some \(\beta\)  and is \emph{closed} if \(d\alpha=0\). Locally closed forms are exact, but globally may not be.

A vector field \(\cX\) and a \(k\)-form \(\omega\)  can be \emph{contracted} to produce a \((k-1)\)-form designated by mathematicians as \(\cX\rfloor\omega\) and given by
\[\cX\rfloor\omega=v^a\omega_{ai_2i_3\cdots i_k} dq^{i_2}\wedge dq^{i_3}\wedge\cdots\wedge dq^{i_k}\]

Both \tm\ and \tsm\ are manifolds in their own right and so one can apply the functors \(T\) and \(T^*\) to these to get higher order iterated bundles such as \(TTT^*T(M)\) or \(T^*T^*(M)\) etc. Four of these are of fundamental importance and will be introduced at the appropriate time.

Let \(\phi:M\to N\) be a map between two manifolds. Given a vector \(\cX\) at \(q\in M\) we can push it to a vector \(\phi_*(\cX)\) (also denoted by \(d\phi(\cX)\)) at the point \(\phi(q)\in N\) as follows: Choose a path \(q(t)\in M\) such that \(q(0)=q\) and \(q'(0)=\cX\), then \(\phi_*(\cX)=(\phi(q(t))'\) at \(t=0\).
One can also pull back a \(1\)-form \(\omega\) at a point \(r\in N\) to a \(1\)-form \(\phi^*(\omega)\) to any point \(q\in M\) such that \(\phi(q)=r\). To define \(\phi^*(\omega)\) it's enough to state what \(\langle\phi^*(\omega), \cX\rangle\) is for any vector \(\cX\) at \(q\). and we define \(\langle\phi^*(\omega), \cX\rangle=\langle\omega,\phi_*(\cX)\rangle \).

\subsubsection{Lifting of vector fields}\label{lvf}
Given a vector field \(\ds\cX=\xi^i\pd{}{q^i}\) on \(M\) there are canonical lifting of it to \tm\ and \tsm. Think of the field as defining an infinitesimal coordinate transformation as \(\tilde q^i=q^i+t\xi^i\) with \(t\) infinitesimal. A vector transforms as \(\ds \tilde v^i=v^j\pd{\tilde q^i}{q^j}=v^i+tv^j\pd{\xi^i}{q^j}\). This means the lifted field in \tm\ is \(\ds \cX_{TM}=\xi^i\pd{}{q^i}+ v^j\pd{\xi^i}{q^j}\pd{}{v^i}\). For \tsm\ a \(1\)-form transforms as \(\ds \tilde \omega_i=\omega_j\pd{ q^j}{\tilde q^i}=\omega_i-t\omega_j\pd{\xi^j}{q^i}\). This means the lifted field in \tsm\ is \(\ds \cX_{T^*M}=\xi^i\pd{}{q^i}- p_j\pd{\xi^j}{q^i}\pd{}{p_i}\).
\subsubsection{Poisson structure}\label{Poiss}
The canonical equivalence \(T^*T^*(M)\simeq TT^*(M)\) exchanges \(1\)-forms and vector fields on \tsm. The vector field corresponding to \(\ds df=\pd{f}{q^i}\,dq^i+ \pd{f}{p_i}\,dp_i\) is \(\ds \cX_f=  -\pd{f}{p_i}\pd{}{q^i}+\pd{f}{q^i}\pd{}{p_i}\). Physicists would call such a function \(f\) a\emph{ hamiltonian} and \(\cX_f\) the\emph{ hamiltonian} vector field. Given two functions \(f\) and \(g\) we define the Poisson bracket by:
\[\{f,g\}=\cX_f(g).\]
A simple calculation reveals the usual formula:
\[\{f,g\}=\pd{f}{q^i}\pd{g}{p_i}-\pd{f}{p_i}\pd{g}{q^i}.\]

The Poisson bracket is anti-symmetric, which the first of the two equations above doesn't show. Another canonical way of defining it is
\[\{f,g\}=\cX_f\rfloor(\cX_g\rfloor d\theta)\]
where the antisymmetry is clear.

\subsection{Geometry of \protect\(\theta\protect\)}\label{aptheta}

\begin{center}
\includegraphics[scale=0.4]{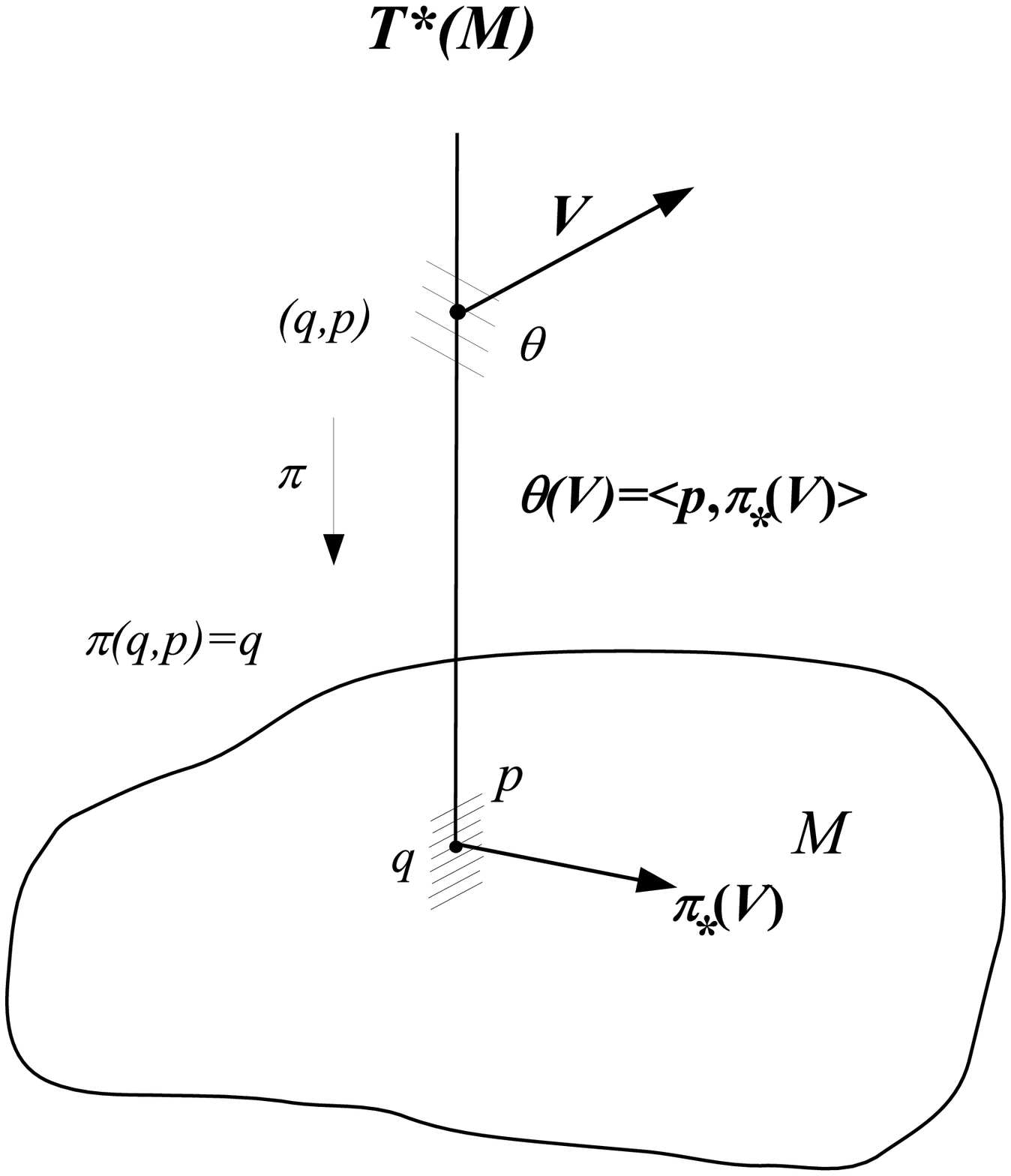}
\end{center}

In this picture \(p\) stands for two things, a \(1\)-form at the point \(q\) in \(M\), explicitly \(p_i\,dq^i\), of which the \(p_i\) are it's components, and also the last \(n\) coordinates of the point \((q,p)\in T^*(M)\). The \(1\)-form \(\theta\) at that point is also written as \(p_i\,dq^i\) where the \(p_i\) are the mentioned coordinates of the point \((q,p)\) used as components of \(\theta\). Of course numerically the \(p_i\) are the same in both usages. In this picture \(1\)-forms are depicted as a series of level curves, actually lines, that is, if \(\alpha\) is a \(1\)-form, then the level surfaces \(\langle \alpha, v\rangle=\hbox{const.}\), on the tangent space of the corresponding point, are a set of lines, and is a graphic way of depicting the \(1\)-form.

The expression \(\langle p,\pi_*(V)\rangle\) is precisely the expression \(p_i\eta^i\) given in the introduction.

One sees that this pictures does nothing more than display  geometrical structures of the cotangent bundle, nothing extraneous to this bundle is brought in. It's remarkable that these structures already pick out a class of dynamical systems, the Hamiltonian ones, which Nature actually employs. This is free lunch.

\subsection{Equivalence of Euler and Hamilton symmetries}\label{eueha}
Let \(X\) be a vector field on \(M\) and \(\ds X_T=\xi^k\pd{}{q^k}+v^j\pd {\xi^k}{q^j}\pd {}{v^k}\) be its lift to \tm\ and        \(\ds X_{T^*}=\xi^i\pd{}{q^i}-p_a\pd{\xi^a}{q^k}\pd{}{p_k}\) its lift to \tsm. Let \(H\) be a Hamiltonian.

For any function \(f(q,p)\), denote by \(f^-\) the function \(f(q,p(q,v))\) where \(p(q,v)=p^-\) is the inverse function to \(\ds \pd Hp\), that is \(\ds\left(\pd Hp\right)^-=v\).

One has by the chain rule
\begin{equation}\label{pchain}
 \pd{f^-}{q^k}=\left(\pd f{q^k}\right)^-+\left(\pd f{p_a}\right)^-\pd{p_a^-}{q^k}
\end{equation}

Suppose \(X_{T^*}(H)=0\), we then have:
\begin{equation}\label{xphz}
 \xi^k\left(\pd H{q^k}\right)^-=p_a^-\pd {\xi^a}{q^k}\left(\pd H{p_k}\right)^-=p_a^-\pd {\xi^a}{q^k}v^k
\end{equation}

One has
\begin{equation}\label{xtl}
  X_T(L)=\left(\xi^k\pd{}{q^k}+v^m\pd{\xi^k}{q^m}\pd{}{v^m}\right)\left(p_a^-v^a-H^-\right)
\end{equation}

Expanding this, lexicographically, we have four contributions

\begin{equation}\label{1-1}
  \xi^k\pd{p_a^-}{q^k}v^a
\end{equation}
\begin{equation}\label{1-2}
 -\xi^k\pd{H^-}{q^k}=-\xi^k\left(\pd H{q^k}\right)^--\xi^kv^a\pd{p_a^-}{q^k}
\end{equation}
\begin{equation}\label{2-1}
  v^m\pd{\xi^k}{q^m}\pd{p_a}{v^k}v^a+v^k\pd{\xi^a}{q^k}p_a
\end{equation}
\begin{equation}\label{2-2}
 -v^k\pd{\xi^a}{q^k}\pd{H^-}{v^a}=-v^k\pd{\xi^a}{q^k}\left(\pd{H}{v^m}\right)^-\pd{p_m^-}{v^a} =-v^k\pd{\xi^a}{q^k}v^m\pd{p_m^-}{v^a}
\end{equation}

Using (\ref{xphz}) in (\ref{1-2}) one sees that all contributions cancel out and we conclude \(X_T(L)=0\).  A similar calculation in the other direction proves the converse.

\section*{Acknowledgements}

My thanks to Harvey Brown for his wealth of ideas and his wisdom.

\end{document}